\documentclass{sig-alternate-05-2015}

\usepackage{balance}

\setcopyright{none}

\makeatletter
\def\@copyrightspace{\relax}
\makeatother

\usepackage{fancyhdr}
\fancypagestyle{plain}{%
   \fancyhf{} %
   \fancyfoot[L]{ACM SIGCOMM Computer Communication Review }%
   \fancyfoot[R]{Volume 47 Issue 2, April 2017}%
}
\def\refname{}

\usepackage{epsfig}
\usepackage{graphicx}
\usepackage{amsmath}
\usepackage{url}
\usepackage{xspace}
\usepackage{enumitem}

\newcommand{\xurl}[1]{{\footnotesize\url{#1}}}
\newcommand{\xref}[1]{\S~\ref{#1}}
\newcommand{\parax}[1]{\noindent \textbf{#1:}}
\newcommand{\proto}{IPIM\xspace}

\newlist{PRIN}{enumerate}{1}
\setlist[PRIN]{label=P\arabic*:,ref=P\arabic*,font=\bfseries,itemindent=\leftmargin\itemsep
0mm\leftmargin 1mm}
\newcommand{\PR}[2]{\item{\textbf{#1:}}\label{pr:#2}}
\newcommand{\PRREF}[1]{\ref{pr:#1}}

\begin{document}

\title{Principles for Measurability in Protocol Design}

\numberofauthors{3}
\author{
\begin{tabular*}{0.99\textwidth}%
{@{\extracolsep{\fill}}ccc}
Mark Allman & Robert Beverly & Brian Trammell\\
\affaddr{International Computer Science Institute} & \affaddr{Naval Postgraduate School}
&        \affaddr{ETH Z\"urich}\\ 
\email{mallman@icir.org} & \email{rbeverly@nps.edu}
& \email{trammell@tik.ee.ethz.ch}
\end{tabular*}  
\begin{tabular}{c}
\end{tabular}\\
\begin{tabular}{c}
\end{tabular}
}

\maketitle

\begin{abstract}

Measurement has become fundamental to the operation of networks and
at-scale services---whether for management, security, diagnostics,
optimization, or simply enhancing our collective understanding of
the Internet as a complex system.  Further, measurements are useful
across points of view---from end hosts to enterprise networks and
data centers to the wide area Internet.  We observe that many
measurements are decoupled from the protocols and applications they
are designed to illuminate.  Worse, current measurement practice
often involves the exploitation of side-effects and unintended
features of the network; or, in other words, the artful piling of
hacks atop one another.  This state of affairs is a direct result of
the relative paucity of diagnostic and measurement capabilities
built into today's network stack.

Given our modern dependence on ubiquitous measurement, we propose
\emph{measurability} as an explicit low-level goal of current
protocol design, and argue that measurements should be available to
all network protocols throughout the stack.  We seek to generalize
the idea of measurement \emph{within} protocols, e.g., the way in
which TCP relies on measurement to drive its end-to-end behavior.
Rhetorically, we pose the question: \textit{what if the stack had
  been built with measurability and diagnostic support in mind?}  We
start from a set of principles for explicit measurability, and
define primitives that, were they supported by the stack, would not
only provide a solid foundation for protocol design going forward,
but also reduce the cost and increase the accuracy of measuring the
network.

\end{abstract}




\section{Introduction}
\label{sec:intro}

The massive size of the Internet and its de-centralized nature make nearly every
facet of its structure and operation complex and often opaque
\cite{clark88design}. Complexity mixed with a lack of visibility is problematic
for not only network diagnostics, but also impacts protocol design and operation
\cite{honda2011}. Further complicating the modern picture is the tension between
measurability and the privacy and security concerns of network operators, users,
and third parties: different parties may view measurability of different
parameters of the network either as essential, or as anathema to best practices.

TCP/IP includes several measurement and diagnostic capabilities that have
evolved in an ad-hoc fashion over time. For instance, ICMP \cite{rfc792} is in
common use to assess reachability and latency, while other facilities are no
longer commonly supported, e.g., IP's Record Route \cite{rfc1812} and timestamp
options \cite{rfc791}. Clever hacks such as \textit{traceroute} and its variants
\cite{Augustin:2006:ATA:1177080.1177100} leverage IP's protection against
infinitely looping packets to elicit router responses that reveal the forward
interface-level data path to a specified target. More convoluted measurements
abound; two such examples are using the IP identification field for alias
resolution \cite{bender2008fixing} and the EDNS-client-subnet extension to map
content distribution networks (CDNs) \cite{calder2013mapping}.

TCP/IP also includes explicit in-band measurement mechanisms, for example TCP's
timestamp option \cite{rfc7323} to assess feedback time and Explicit Congestion
Notification (ECN) \cite{rfc3168} to allow routers to signal congestion to end
hosts. Despite these intended (and unintended) protocol hooks, the
measurability and accountability of the network has historically been a
secondary design concern~\cite{clark88design}. Retrospectively, the diagnostic
facilities currently available have proven woefully inadequate for applications,
operators, policy makers, and researchers on the modern Internet in several
ways:

\begin{itemize}
\itemsep 0pt
\item The diagnostics built into TCP/IP are useful for measuring a few specific
  attributes of the network, but are not germane to the breadth of understanding
  we now desire. For instance, there is no ready way to explicitly understand
  available capacity or packet manipulation along some path, both of which are
  now more important than when TCP/IP was developed.

\item IP addresses are no longer useful as host identifiers for many measurement
  purposes. Today's Internet has both multiple machines sharing an IP address
  (e.g., NATs, anycast) and single machines using multiple IP addresses (e.g.,
  routers, load-balancers). An interface may have both IPv4 and IPv6
  identifiers, and, in the case of IPv6 privacy extensions, each IPv6 address
  may be used only once or for a short time \cite{plonka2015temporal}. Beyond
  network layer addresses, CDNs hide myriad hosts behind a common name.  Even
  resolving a hostname often involves a mysterious chain of resolvers
  \cite{weaver2011implications,SCRA14}.

\item Today's network forwards different types of traffic with different
  policies.  Therefore, a measurement leveraging a diagnostic like the ICMP echo
  mechanism may not be indicative of the experience of a web transfer.  This
  situation renders even experienced operators at a loss when problems arise.

\item Protocols and systems are increasingly reactive to network conditions.
  For instance, mobile devices may prefer WiFi networks over cellular networks
  when both are available.  Or, CDNs attempt to direct clients to the ``best''
  content replica.  Or, the Happy Eyeballs technique \cite{rfc6555} is used for
  assessing the quality of IPv6 versus IPv4 to reach a destination. While it is
  perfectly reasonable to base operation on network conditions, obtaining such
  understanding is an ad hoc and arduous process that is generally performed
  out-of-band without any unified mechanism.

\end{itemize}

Over the years researchers---ourselves included!---have designed
increasingly complex and clever methodologies for understanding the operation of
the Internet. This large body of work unquestionably provides many insights into
current Internet reality.  However, we find three issues with the current
state-of-the-art:

\begin{itemize}
\itemsep 0pt
\item Often techniques rely on inference and not direct measurement.  That is,
  we are forced to make assumptions of varying dubiousness about the network, or
  how traffic is handled by the network \cite{paxson2004strategies}. For
  instance, that IP addresses represent unique machines. Or, that hosts generate
  TCP acknowledgments quickly. Or, that a vantage point ``close'' to an endpoint
  is ``good enough'' to well understand the endpoint's perspective.  Even more
  problematic, measurements face an increasingly adversarial environment where
  honeypots and deception pollute our understanding \cite{alt2014uncovering},
  while increased deployment of encryption \cite{naylor14} can remove visibility
  completely.

\item We often use complex post-facto analysis to illuminate various network
  behaviors.  This leaves us with a general understanding of some phenomenon,
  but not a way to generate such understanding on-the-fly such that it
  constitutes actionable information within some system.

\item The techniques we employ often exploit unintentional behavior in order to
  coax information from the network that operators may not wish to divulge.
  Simultaneously, operators are unable or unwilling to deploy or use many of the
  techniques developed in the research community.

\end{itemize}

In this position paper, we describe the In-Protocol Internet Measurement
(\proto) facility.  Rather than the point solutions found in individual
protocols or brittle tricks employed by researchers, we seek to identify a
minimum set of measurement \emph{primitives} that generalize across a wide
variety of use cases.  \proto is a proposal to take these primitives and promote
measurement to a first class citizen within the network architecture.   

We posit that by adding explicit measurement primitives into the protocol stack,
a broader and more accurate understanding of network behavior will be available
to the protocols themselves, applications, operators, developers, and
researchers.  Further, this understanding will come at a lower cost than
attempting to leverage the current accumulation of measurement hacks and
assumptions.

\section{A Motivating Example}
\label{sec:ex}

As a brief motivating example we consider the difficulty surrounding a seemingly
simple task: round-trip time (RTT) assessment. Latency is one of the fundamental
properties of network paths, having an impact on everything from when to
retransmit a packet within a reliable stream \cite{PACS11} to protocol
performance \cite{padhye98} to determining the magnitude and amount of network
congestion \cite{luckie14}.  Indirectly, latency is also used for geolocation
and to direct queries among content caches.

RTTs can be measured out-of-band using ICMP's echo facility.  While in some
cases ICMP can provide acceptable answers, it also presents three drawbacks:
($i$) ICMP is often blocked as a matter of policy, rendering this technique
useless, ($ii$) leveraging ICMP makes an assumption that the network treats ICMP
the same as other more user-oriented traffic such as TCP/HTTP and therefore that
the RTTs from ICMP measurements are germane for other traffic types and ($iii$)
when an out-of-band ICMP mechanism is used within a larger system or protocol it
represents additional complexity to develop and maintain.  

An alternative approach is to utilize natural protocol interactions to measure
the RTT.  For instance, we could leverage small DNS requests and responses or
TCP data segments and the corresponding acknowledgments (ACKs).  The problem is
that these interactions often include more than the network latency.  For
instance, if a DNS request arrives at a resolver that does not have the
requested name in local cache, then there will be additional time in iterating
through the DNS hierarchy (or some subset thereof) to obtain the answer to the
query.  However, the contents of the DNS response will be nearly identical
regardless of the state of the cache, leaving the requester blind to whether the
RTT reflects the network path to the resolver or not.  Of course, the astute
Internet empiricist will no doubt decide to pile on another hack and send the
same DNS request twice, using on the second request and its corresponding
response as the network RTT on the assumption that the first request will prime
the resolver's cache---an assumption which may or may not be true. However, even
in the best case where this second transaction is an accurate assessment of the
network path, the process reverts back to out-of-band measurement since the
naturally occurring DNS transaction is insufficient on its own.  The same
ambiguity plagues web-based RTT measurements, where web content is today
commonly cached or proxied.

TCP is a canonical example of a protocol whose messages allow it to infer path
properties (including RTT estimation). However, TCP must contend with a variety
of complicating factors including delayed ACKs \cite{APB09}, hardware
acceleration (e.g., offload engines), and retransmitted segments. These can skew
RTT estimations by hundreds of milliseconds---an error that can be of the same
magnitude as the actual RTT. TCP's timestamp option \cite{rfc7323}---an explicit
timestamp included by the sender in each segment and echoed back by the
recipient---helps resolve the retransmission ambiguity, and we utilize a similar
mechanism in \proto to decouple host and network latency.

While in-band measurement is powerful, DNS and TCP well-illustrate the
difficulties and subtleties in obtaining accurate measurements via natural
interactions.  \proto introduces a set of primitives that provide more precise
and less inferential measurements, while increasing the space of measurable
network properties.  These primitives are designed to generalize across a wide
array of protocol, application, and user needs (RTT estimation is but one
use-case; \xref{sec:use} details additional application of \proto). Improvements
to \proto can then benefit a wider range of protocols, and free designers from
creating yet more point solutions.

\section{Principles}
\label{sec:prin}

Although the original ARPANET contained extensive explicit support for
measurement and diagnostics \cite{CrovellaKrishnamurthy}, measurement in the
Internet has long been viewed as a management function decoupled from protocol
design.

In our experience running and measuring networks, creating protocols, and
observing the evolution of deployed protocols and systems, several common themes
among successful designs have emerged. From these design trends we extract a set
of principles to guide our design of \proto:

\begin{PRIN}

\PR{Measurement should be explicit}{explicit} This principle is not unique to
Internet measurement~\cite{pep20}, but is especially applicable to it. Many of
the techniques applied in Internet measurement rely on inference that is itself
based on assumptions about the reactions of protocol implementations outside the
control of the party performing the measurement, assumptions that may or may not
hold. For example, stretch ACKs in TCP~\cite{APB09} invalidate basic assumptions
in passive latency measurement, and differences between the addresses of
interfaces on which a packet is sent and the addresses reported by routers in
ICMP Time Exceeded replies complicates the analysis of traceroute results when
studying data-plane topology. The only way to address this problem is with
facilities for measurement which state the assumptions on which inferences are
to be based explicitly. Measurement facility explicitness also encourages
adoption, rather than the wholesale blocking of measurement traffic when
administrators do not clearly understand what is being measured and what
information is being shared.  In \PRREF{control}, we elaborate on providing
hooks for user and administrator consent.

\PR{Measurement should be in-band}{inband} Modern Internet paths represent a
complex forwarding fabric whose behavior is driven by a multitude of properties,
including traffic type and content, policies, load, and the end systems
involved. This makes out-of-band measurements inherently tricky.  Often at least
one of the properties the path uses to make forwarding decisions will differ
between operational transactions and synthetic measurement traffic---leading to
the measurement not faithfully capturing the path's true treatment of particular
traffic.  In turn, this impacts the insights we gain from such measurements in
unknown ways.  Thus, measurement \emph{within} a protocol is not the same as
out-of-band measurement.  Therefore, faithfully understanding production traffic
calls for in-protocol primitives.

\PR{Measurement consumer bears cost}{cost} Designing protocols to enable
measurement and introspection inherently imposes compute and memory
requirements.  We strive for designs that minimize state and per-packet
processing, especially within the core of the network. Our goal is to collect
myriad small and simple bits of information provided by end systems and routers
to gain broad understanding about the Internet.  We then aim to concentrate the
costs of understanding these bits of information with the actor interested in
the measurement to the extent possible.  As an example, data collection can be
probabilistic such that measurement data is contained within only a subset of
packets or flows.  Situations requiring more granular, precise, or
representative measurements can in turn employ broader or longer data
acquisition. More generally, designing probabilistic behavior into measurement
primitives allows them to be deployed with an explicit trade off between
accuracy and overhead.  This need not, however, limit the accuracy or coverage
of the measurements available, by shifting the balance of effort from the
runtime measurement process to a post-runtime analysis process.

\PR{Measurement provider retains control}{control} Within the tussle space of
measurement, operations, and security, external measurements are frequently
viewed as intrusive, violations of policy or privacy, or simply unwanted.
Application developers, network operators, and end users must have control over
how much information is sent to the peer and/or exposed to observers along the
path. Control over measurement at the endpoints also allows measurement features
to be selectively enabled in order to help diagnose issues for specific traffic
flows.

\PR{Measurement must be visible}{visible} The widespread deployment of
asymmetric routing implies different behavior on the forward and reverse paths,
and the opacity of one or another of these paths to one-way measurements limits
the insight such measurements can provide. Therefore, the ability of a packet
recipient to echo information back to the packet source is crucial to increasing
visibility into the path.  The notion of visibility extends to
\PRREF{control}, above. And, while integrity over measurements is essential,
particularly in a world of network devices that transparently
intercede, encryption or data obfuscation should not be
relied upon as an enabler of measurements.  More fundamentally,
network elements that cannot understand what measurements are being
made may universally block all such measurements. 

\PR{Measurement should be cooperative}{cooperation} The Internet has diverged
from a clean end-to-end model. Routers and middleboxes on path actively
manipulate the data plane, and this design pattern can be extended to improve
measurement. Routers and middleboxes should participate in measurement not just
via their control plane, but also via inspection and marking of the data plane.

\end{PRIN}

\section{Primitives}
\label{sec:prim}

Having distilled the principles underlying the design of \proto, we now turn to
sketching \proto's information model. In this paper our goal is to sketch
\proto at a high level. Therefore, we do not focus on details such as header
layout, information granularity, or counter sizes: a full protocol
specification is left to future work.

Prior proposals, e.g., \cite{ipmp}, and standards, e.g., \cite{rfc4656}, define
new out-of-band protocols and mechanisms for path and delay diagnostics akin to
a more featureful ICMP.     Instead, our vision is broader---preferring to not
only assess per-hop characteristics, but allow for measurements to be taken as a
side-effect of normal protocol interactions, and hence serve as actionable input
into protocol, network, and application operation.  

While we phrase these primitives in terms of information the sender exposes to
the receiver to allow both sides to make measurements, we note that all of this
information is available to passive observers as well. \proto can thus be used
to expose information which can be passively measured in aggregate. For example,
a network border monitor can observe end-to-end timing information
(\xref{sec:prim:e2e:time}) to detect and react to path changes or congestion via
analysis of the time series of latency among network pairs.

Note also that it is not necessary to provide all the information described in
this section in each packet, or even within each flow. Different uses of \proto
will require different types of information at varying granularity.  Therefore,
\proto's use can be shaped for each specific need.  For example, flows could be
selected with a given probability to contain arrival information and hence
provide a general characterization of the path without involving all
connections.

Finally, we do not claim this sketch is complete or optimal. Instead \proto
is extensible such that additional information can be included in the future
as needs arise.

We start in \xref{sec:prim:e2e} by describing two primitives that operate only
end-to-end; i.e., that require no interaction with devices along the path and
can therefore be implemented and deployed today. We then describe primitives
that require cooperation from devices along the path. We consider primitives
that only require assistance from a subset of the intermediate devices in
\xref{sec:prim:hop} and then move on to primitives that require cooperation of
more or less all devices along a given path in \xref{sec:prim:acc}. While more
difficult to deploy, these latter primitives illustrate what is possible when
including measurement as a first-class citizen within the protocol stack.

\subsection{End-to-End Information}
\label{sec:prim:e2e}

We start with information that can be collected with only the participation of
the end systems.

\subsubsection{Host Identification}
\label{sec:prim:e2e:hostid}

Increasingly, IP addresses do not identify hosts.  Load balancers, anycast,
IPv6, NAT, and the like frequently confound efforts that assume a
host-to-address bijection. To calibrate the measurements \proto enables, we aim
to understand precisely which host is involved in some communication by defining
a host identification field $H_{id}$.  Host identifiers have been explored
extensively in the past; see for instance literature on divorcing host
identification from location of attachment, e.g.
\cite{Carpenter:2014:IAC:2602204.2602215} and references within.  Whereas this
prior work uses such identifiers for discovery and delivery, our use is limited
to measurements and not for traffic routing.

In many cases stable host identifiers are not problematic, such as when
identifying some replica of a service.  However, in other cases they could help
track users across time and attachment points. To balance measurement goals with
privacy requirements, we offer two techniques to thwart tracking. First,
$H_{id}$ will be modest in size---e.g., 16~bits---to disambiguate multiple
machines behind some common IP address.  That is, $H_{id}$ is not meant to be
globally unique. By the pigeon-hole principle, globally, or even behind large
NATs, there will be many hosts sharing the same $H_{id}$ and so a given $H_{id}$
cannot be used to track a given host (or user) over time. Second, since this
identifier is meant for alias resolution, and not part of the data delivery
process, it can be changed often and according to an end host's own policies
without external coordination. For instance, while $H_{id}$ must be stable
across connections and transactions to be useful, $H_{id}$ could be randomly
reset every 30~minutes and/or each time the host joins a new network or obtains
a new dynamic IP address. Intentional identifier collisions combined with
regular random identifier rotation enables measurements to disambiguate hosts,
while still adhering to the security and privacy principle \PRREF{control}.  In
cases where the population under study is large and induces many identifier
collisions, multiple observations over time suffice to probabilistically
disambiguate by principle \PRREF{cost}.

\subsubsection{Timing Information}
\label{sec:prim:e2e:time}

TCP's timestamp option \cite{rfc7323} points to a way to leverage information
within a protocol to understand latency. TCP senders include both the current
timestamp and the most recent timestamp received from the peer when sending a
segment.  In this way, a host can compare the echoed timestamp with the current
time to determine the feedback time upon each packet arrival. Beyond TCP, \proto
provides a general mechanism for protocols, applications, and passive observers,
to utilize latency information.

As discussed in \xref{sec:ex}, TCP's mechanism suffices for use in determining
the retransmission timeout (RTO), but for measuring network properties a key
deficiency is that the latency measurements can include a non-trivial amount of
non-network induced time introduced by TCP's delayed ACK mechanism \cite{APB09}.
Therefore, simply taking the difference of the current time and the echoed time
includes not only network latency, but also host-based delay.

\proto generalizes and improves upon TCP's approach by using a timing tuple
$(T_{now},T_{echo},T_\Delta)$, where $T_{now}$ is the time the given segment was
transmitted, $T_{echo}$ is the timestamp from the previous packet to arrive in
this flow (connection) and $T_\Delta$ is the interval between when the previous
segment was received and the current segment is being transmitted.  This gives
insight into the fraction of the round-trip delay added by the end hosts.  For
instance, consider a data segment sent by a host with $T_{now}=45$ and a
corresponding ACK received at time 95 with a timing tuple of $(143,45,15)$. This
tells us ($i$) the ACK was sent at time 143 according to the receiver's clock,
($ii$) the full amount of time between sending the data segment and receiving
the ACK is $95-T_{echo}=95-45=50$, and ($iii$) the amount of network-based delay
in this case is $95-T_{echo}-T_\Delta=95-45-15=35$. Note that, following
principle \PRREF{cost}, this calculation requires only a small amount of state
to be kept at either endpoint (i.e., the arrival time and $T_{now}$ from the
last packet arrival). Finally, in addition to raw latency, this timing tuple can
have additional uses---e.g., to understand how packets are spread or compressed
as they traverse the path for the purposes of capacity assessment
\cite{dovrolis03}.

\subsubsection{Arrival Information}
\label{sec:prim:e2e:arr}

Given the Internet's ``best effort'' nature, the source of a stream of packets
is given no information about their handling by the network.  Some reliable
protocols like TCP build machinery into the protocol that gives the sender some
understanding of how the path is behaving---e.g., some notion of the packet loss
rate and/or packet re-ordering.  TCP gives the sender a rough understanding of
these path properties by using the acknowledgment stream as a crude form of the
echo facility we describe in \PRREF{visible} (\xref{sec:prin}).  The ACKs give
the sender a view, albeit limited, into the arrival process at the receiver.
This same viewpoint is not naturally available for unreliable traffic or traffic
for which reliability comes from heavy coding of the packet stream such that
receiver could fix losses without involving the sender.  In these cases, the
sender remains in the dark about the path's treatment of the packet stream.  In
\proto we seek to better illuminate path behavior \emph{across} protocols and
applications.

To understand arrival patterns, \proto includes a nonce tuple \\
$(N_{xmit},N_{sum})$ roughly modeled after the ``cumulative nonce'' from
\cite{savage99a}.  The sender populates $N_{xmit}$ with a random but increasing
value before transmitting the packet.  The receiver simply sums (modulo the size
of the field) the incoming $N_{xmit}$ values and places that value in the
$N_{sum}$ field of its outgoing packets.\footnote{This mechanism differs from
\cite{savage99a} in that only in-sequence segments contribute to the nonce sum
in the previous work.}  In this way the sender can reconstruct the arrival
stream and pattern.  For instance, consider the case when a sender transmits
three segments with $N_{xmit}$ values of 5, 1001 and 5800 and receives two ACKs
in return with tuples (45,5800) and (1376,5805).\footnote{These values are
chosen to be expository; a real system would maximize information gain via
coding.}  From the ACKs the sender can conclude three things: ($i$) the ACKs
were not mis-ordered because the $N_{xmit}$ values from the receiver are
increasing (from 45 to 1376), ($ii$) the second segment sent by the sender (with
$N_{xmit}=1001$) was lost since it is not included in the $N_{sum}$\footnote{It
is possible that the second segment was mis-ordered and will be included in a
future $N_{sum}$.  The sender will have to wait for a long enough period of time
to disambiguate loss and reordering.} and ($iii$) the first and third packets
arrived out of order since the $N_{sum}$ reflects a packet with $N_{xmit}=5800$
before including a packet with $N_{xmit}=5$.  We note that this mechanism does
require state and processing to understand the path properties. However, per
\PRREF{cost} in \xref{sec:prin} we lay this burden on the endpoint deriving the
information.\footnote{Also, note, that if the sender uses a progression of
values for $N_{xmit}$ that is not strictly random, but uses some pattern of the
source's choosing then understanding $N_{sum}$ values that reflect loss and
reordering may be easier than just using strictly random values and then using
brute force to reconstruct the events.} A recipient need only keep the current
nonce sum and perform an add on each packet arrival. Finally, we note that this
mechanism also suffices for identifying misbehaving receivers that are trying to
coax the sender into transmitting faster than normally allowed, which is the
original intent of the cumulative nonce in \cite{savage99a}.

\subsubsection{Integrity Information}
\label{sec:prim:e2e:int}

A final piece of \proto's end-to-end information provides an integrity check
over the outgoing packet per principle \PRREF{cooperation}. We take inspiration
from HICCUPS \cite{CBA14b} and include a basic integrity tuple
$(I_{cover},I_{mode},I_{hash},I_{echo})$, where $I_{cover}$ is an indication of
which parts of the packet are covered by the integrity check; $I_{mode}$ is an
indication of which mode the integrity check is using (see below); $I_{hash}$ is
a computed hash across the covered fields of the packet; and $I_{echo}$ is the
most recent $I_{hash}$ to arrive.  The integrity check can use several modes.
First, the sender can compute $I_{hash}$ using only a well-known hash function
and information within the packet itself.  This allows the receiver to
understand packet manipulations, but also would allow a middlebox to trivially
re-compute the hash in a manipulated packet.  A second mode calls for including
a salt that only the sender knows in the $I_{hash}$ value.  While this mode
prevents middleboxes from re-computing the $I_{hash}$, it also prevents the
receiver from directly understanding manipulations.  A final mode calls for
computing $I_{hash}$ with a salt known to both endpoints (and arranged
out-of-band).  While this allows the receiver to understand manipulations while
also preventing trivial $I_{hash}$ replacement by middleboxes, it requires a
shared secret between the endpoints.

Note that $I_{cover}$ can vary over the course of a transaction. While
$I_{echo}$ can only give a binary indication of whether the given packet fields
have changed, across a natural flow of packets with varying coverages, we can
determine which parts of the packets are being transformed by the network.

\subsection{Hop-Specific Information}
\label{sec:prim:hop}

While end hosts can usefully illuminate many aggregate path properties without
directly involving the intermediate hops, we can often get additional or better
information by directly engaging these network elements (principle
\PRREF{cooperation}).  Adding a requirement that all routers and in-path devices
process all packets to assist with measurements is clearly burdensome.
Therefore, before discussing the types of information routers can provide, we
introduce two strategies for collecting hop-specific information, as follows.

\parax{Probabilistic Stamping} Using this strategy, a router first
samples a small fraction of the packets it forwards (principle
\PRREF{cost}).  If a sampled packet includes an \proto request for information,
the router fills in the requested information before forwarding the packet (see
below for specifics about the information that can be requested).  The
information from the router is echoed back to the source by the final recipient.
The router includes the current IP TTL in the stamped information, which allows
the sender to understand the relative location of the hop providing the
information.   The sampling rate---which could vary depending on the load on the
router---allows routers to directly control the burden imposed by these
measurements, or opt-out altogether (principle \PRREF{cost}).

\parax{Triggered Stamping} We envision probabilistic stamping to be useful in
developing a general understanding of the path and its properties.  However, we
introduce an additional technique to obtain information about a specific hop:
triggered stamping.  In this scenario a target TTL is given by the packet source
and when the IP TTL equals the target TTL a router stamps the packet with the
requested information.   As above, the ultimate packet destination would echo
the information back to the source.  This closely follows the spirit of the ICMP
Time Exceeded message which calls for router action when the IP TTL reaches
zero.  \proto simply extends this to triggering action on a particular TTL value
and the information is then included in standard transactions rather than
out-of-band in an ICMP message.

Note, for either the probabilistic or triggered strategy, we follow the ``best
effort'' principle and do not require routers to participate.  We aim to make
participation fairly low cost, but it will clearly be greater than simply
forwarding a packet.  When measuring the system under heavy load a tussle
arises.  On the one hand, this is a point where understanding and therefore
measuring the system is particularly crucial.  On the other hand, we do not wish
to exacerbate operational problems for users.  Therefore, we advocate a system
where each router can manage its own resources and decide what to provide and
the end systems must cope with incomplete information. Finally we note that
while full deployment by all hops is ideal, the design admits incremental
deployment and, where deployed, aids understanding of hop-specific details of
the network.

The per-hop information \proto stamps in packets can take several forms.  Here
we sketch two tuples, although more could be added.

\begin{description}
\itemsep 0pt
\item [Topology Information:] The first set of information routers may stamp
  relates to topology.  We define the following topology tuple:
  $(ID,AS,IP_A,IP_D)$, where $ID$ is a unique ID for the given router assigned
  by the router owner, the $AS$ is the autonomous system in which the router
  resides and $IP_A$ and $IP_D$ are the IP addresses of the interface the packet
  arrives and departs on, respectively.
  \vspace{3mm}

\item [Performance Information:] Routers may include performance information by
  encoding the tuple: $(T_{now},QL,AC,CL)$ where $T_{now}$ is the current time,
  $QL$ is the queue length expressed as time,\footnote{Routers often have
  multiple queues that packets traverse and $QL$ should represent the aggregate
  queuing time for the hop.} $AC$ is the available capacity,\footnote{Encoded in
  coarse terms, a la Quick Start \cite{FAJS07}.} and $CL$ is the current
  congestion level---which is meant to be a finer grained version of ECN's
  binary feedback.
  \vspace{3mm}
\end{description}

In addition to the hop-specific information \proto gains each time a router
stamps a packet, the information from multiple packets can be combined to form
broader understanding.  Over time, the \proto stamped traffic will provide the
IP- and AS-level path over which the communication traverses.  Traditional
topology ambiguities, such as the AS to which a router belongs or the set of
interface aliases, become explicit per \PRREF{explicit}.  Further, topology can
be combined with performance information to more deeply understand how
individual hops contribute to overall path properties, as opposed to our current
situation of attempting to infer such information from suboptimal and sometimes
dubious sources of information \cite{willinger2009mathematics}.

A final note is that by providing space in a packet for a single router to
populate, we side-step the fragmentation problem that results if routers are
permitted to append information to the packet.  That is, if each router were to
add topology information (a la IPMP \cite{ipmp}) the packet would grow to the
point of requiring fragmentation (or carry little, if any, actual payload data).
By allocating fields in the packet for this information at transmission time we
do not fall prey to this problem.

\subsection{Accumulated Path Information}
\label{sec:prim:acc}

Hop-specific measurements (\xref{sec:prim:hop}) are designed to develop an
understanding of the specific hops that make up the path over some period of
time (and traffic), not instantaneously.  We now add a third category of
information that is not easily developed using either or both of the strategies
from the previous two subsections: accumulated path information.  That is,
information that spans multiple hops, but cannot be understood without help from
the routers.  As with the rest of \proto, the set of accumulated path properties
we measure is extensible, but in this initial work we consider two sets of
information, as follows.

\begin{description}
\itemsep 0pt
\item [Path Changes:] The first set of information aims to give an indication
  when the path between two endpoints changes.  For instance, this could inform
  a congestion control scheme that it's understanding of the network path is
  out-of-date and needs to be re-learned. To understand path changes we
  introduce an evolution tuple $(E_{cur},E_{echo})$ that works similarly to the
  TTL.  The sender initializes $E_{cur}$ to some random value.  Each router $R$
  chooses some long-lived random offset $O_R$ and adds $O_R$ to $E_{cur}$ in
  each packet before forwarding, wrapping around on overflow. $O_R$ can be
  positive or negative, and can be large relative to the range of $E$. The
  receiver simply echos the value back in $E_{echo}$ field.  The difference
  between the starting $E_{cur}$ and the received $E_{echo}$ at either endpoint
  should be constant (regardless of exact starting point for $E_{cur}$) for a
  path that does not change.  When a path change causes a different set of
  participating routers to be seen, their different $O_R$ values will alter the
  difference the end host observes.

\item [Performance Information:] We revisit the performance information we
  develop in a hop-specific manner in \xref{sec:prim:hop}.  Instead of focusing
  on a single hop at a time, it will sometimes be useful to understand
  accumulated state of all hops at about the same time.  As an example, previous
  work in Quick Start seeks to have routers validate an initial sending rate for
  a flow \cite{FAJS07}.  This means that each router must consider the target
  sending rate in a packet and, if the rate is deemed too high, lower the rate
  before forwarding the packet.  With this in mind we define a performance tuple
  $(AC_{min},QL_{sum})$ where $AC_{min}$ is the minimum amount of available
  capacity at any hop along the path (as in the last subsection, this could be
  coarsely encoded, a la Quick Start) and $QL_{sum}$ is the total amount of
  queuing delay due to participating routers.
\end{description}

Since the accuracy of accumulated path information increases with the proportion
of routers participating, there are two additional considerations.  First, it
should be (relatively) inexpensive for a hop to contribute information.  In
particular, the information added to the traffic should remain static over some
period of time and not require per-packet analysis.  Further, as we sketch
above, this should be considered as ``best effort'' information to which the
router can choose not to contribute when under high load.  Finally, since we aim
eventually to accumulate over all hops in the path, we need some way to
understand whether all hops did indeed contribute.  For this we borrow the Quick
Start strategy of using a TTL$'$ field which starts at some random value and
yields a $TTL_\Delta = |TTL-TTL'|$. The TTL$'$ is decremented by one by each
router that considers the accumulated information request.  The destination of
the packet echos back the absolute difference between the TTL and TTL$'$. If the
echoed difference is equal to the initial $TTL_\Delta$ then all routers
contributed to the accumulated path information.  This mechanism helps to
explicitly calibrate the information that comes back from the network so we do
not have to guess how much of the path contributes to the measurement
effort.

\section{Use Cases}
\label{sec:use}

We next discuss a number of \proto use cases.  These are not meant
to be exhaustive, but rather are illustrative of the breadth of the
utility provided by the facility.

\vspace{0.6mm}
\parax{Protocol Adaptation}
Applications, systems and the underlying transport protocols often base their
activities on some understanding of the network path. Canonical examples include
TCP's algorithms for adapting the sending rate to the network path's current
congestion level and CDNs routing application traffic to the best replica when
necessary.  \proto helps these cases in two important ways: ($i$) the assessment
of the network becomes a standard facility instead of something that must be
designed and built for each use, which reduces the cost of each use of the
information and ($ii$) the granularity and accuracy of the information can be
better (e.g., congestion control reductions in the sending rate can be tailored
to the available capacity reported instead of blindly halving the rate).

\vspace{0.6mm}
\parax{Network Mapping}
In a network in which most routers support \proto, hop-specific information
(\xref{sec:prim:hop}) provides the possibility to replace \textit{traceroute}
with explicit topology and network structure discovery without
\textit{traceroute}'s perennial drawbacks \cite{willinger2009mathematics}:
potential differences in treatment between traceroute packets and production
traffic, aliasing, and lack of information about the reverse path.  In
particular, routers that add \proto topology stamps obviate the current
expensive and error-prone requirement to reduce IP interfaces returned by
traceroute to routers (the aliasing problem). Such topology information is
valuable to enterprises for management and debugging, and invaluable to large
providers seeking to optimize content delivery speed and reliability.

\vspace{0.6mm}
\parax{Management}
Many network management tasks ultimately involve understanding how a network is
operating at the current time, determining whether the current performance is
problematic and, if so, taking steps to fix the underlying issues. \proto can
assist the first two steps of this process.  A network that can monitor the
performance-related measurements---e.g., latency---carried on production traffic
can develop a notion of the overall state of the network as perceived by normal
traffic.  This can in turn be used to develop both a baseline performance
expectation and therefore deviations from this baseline.

\vspace{0.6mm}
\parax{Informing Policy} 
The role of regulation and policy in shaping the Internet has
steadily increased.  Concerns over network neutrality
\cite{kreibich2010netalyzr},
reliability \cite{Quan:2013:TUI:2486001.2486017}, consumer choice, and
competition \cite{sundaresan2011helping} have
risen to the forefront as the Internet as become critical
infrastructure.  Policy and regulation are best shaped by empirical
observation and therefore \proto is particularly apropos in this
space as it offers the ability to concretely expose and attribute network
structure, behavior, and the treatment of traffic.

\vspace{0.6mm}
\parax{Data Centers}
Large amounts of content and computing are currently handled by data
centers, which are often self-contained homogeneous networks in
their own right.  These environments may be early adopters of \proto
as a data center often has stringent performance requirements and is
under a single administrative control structure, allowing \proto to
be more easily deployed.  \proto can expose timing, capacity, and
delay information about production traffic flows in order to quickly
adapt to real-time network conditions, load, and faults.

\section{Practicalities}
\label{sec:practical}

While our goal in this work is to explore an architecture that generalizes
Internet measurements into a cohesive facility, we must be guided in part by
eventual deployment constraints. Here, we discuss several practical
considerations for \proto.

\subsection{Overhead}
\label{sec:practical:overhead}

First, we consider \proto's overhead. \proto's new functionality comes at an
inevitable cost. We consider two aspects of this cost in this initial work:
($i$) transmission overhead (bits on the wire required in each packet) and
($ii$) the processing costs in the devices that provide and process \proto
information.

\vspace{3mm}
\parax{Packet Overhead}
At first glance, adding measurement information inline would seem to impose a
prohibitive per-packet overhead. However, three aspects of \proto's design
reduce this overhead to a modest per-packet cost. First, not every
communication will care about every network property \proto can assess. E.g.,
an application may wish to understand latency and topology, but not available
bandwidth or middlebox interference. Therefore, at most only a subset of
\proto's capabilities will be brought to bear within each flow. Second, periodic
samples will often suffice, eliminating the requirement to place measurement
information on every packet. In line with \PRREF{control}, originators can
provide or request information be sampled per packet or per flow, depending on
their measurement requirements. This also means that the measurements can be
spread out such that each packet contains a single kind of measurement.
Finally, compact encoding of \proto information can reduce overhead on those
packets containing measurement information while shifting the cost of processing
to the measurement consumer, in line with \PRREF{cost}. For instance, consider
the timing tuple we introduce in \xref{sec:prim:e2e:time}:
$(T_{now},T_{echo},T_\Delta)$. This could be encoded in a 32~bit field with each
timestamp being allocated ten bits and the two remaining bits being used to
indicate the granularity of the timestamps.  This would suffice for the vast
majority of latency sampling tasks at a cost of less than 0.3\% of the space of
a typical 1500~byte packet.

\vspace{3mm}
\parax{Processing Overhead}
As we sketch in \PRREF{cost} in \xref{sec:prin}, one of our design guidelines is
that ($i$) most actors in the system should provide simple information that does
not require significant new state and ($ii$) the beneficiary of the information
bears the burden of analyzing the measurements to obtain high-level insight.
This is especially important when information is collected from intermediate
nodes---switches or routers.  Therefore, our design often calls for ``stamping''
packets with already known information instead of computing some new information
and including that on the passing traffic. This packet manipulation is more akin
to ECN marking of packets than, for instance, generating an entire new packet
(\`{a} la ICMP echo response or time-exceeded message, which also impose higher
transmission costs). Further, we explicitly consider the information ``best
effort'' in that if a node is resource constrained, \proto will not pose further
cost. Finally, we design for nodes to act on information in a probabilistic
fashion so that nodes can ultimately enact policy that bests suits their needs
as they balance the load imposed by \proto with the node's myriad other tasks.

\vspace{3mm}
\subsection{Information Location}
\label{sec:practical:loc}

There is a natural question of where to place the \proto information within
packets. The principles and primitives we describe stand without regard to where
the information is placed. However, \PRREF{visible} calls for \proto information
to be visible to passive observation, so it needs to be easy for observers to
know whether a given packet contains \proto information and where it is located.
Further, if in-path nodes such as switches and routers are to effectively
participate in \proto, this information must appear at a fixed, or at least
constant-time computable, location within the packet.

Ideally, we would build a thin measurement layer between the network and
transport layers. This is a clean approach that satisfies the above criteria.
However, wedging a new layer into the protocol stack may be too high of a burden
for existing stacks and network devices. Another possible location is building
\proto into the network layer, via extension headers within IPv6 (or IP options
within IPv4), though the deployability of such headers is
questionable~\cite{RFC7872}. Understanding the pros and cons of these two
approaches will be one of the crucial first tasks as we move from a conceptual
design to an instantiated system.

Regardless of the location of the information, we note that the originator will
need to build space into packets to carry any information provided by in-path
devices. This is an important consideration as increasing packet sizes during
transit may require fragmentation which would make \proto practically
unworkable.

\subsection{Incentives}
\label{sec:practical:incent}

The current ad-hoc use of inferences to drive the operation of the Internet
illustrates that measurements are crucial and the effort to obtain information
about operational networks is significant. Hence, the benefits of
\proto are clear.  As detailed in our use cases (\xref{sec:use}), \proto
benefits not only network operators, but also researchers, policy makers,
applications, and, ultimately, users. However, \proto still must overcome not
only the inertia facing any new technology, but must present a compelling
advantage.

Instrumentation and measurement is now common in data centers, among providers,
and even within enterprises to increase performance, reliability, and
utilization---all of which are driven by strong economic incentives. \proto
would provide a standard mechanism for all nodes, applications, and users to
perform detailed introspection for their own benefits.  And while some parties
may be naturally adverse to exposing any information, \proto's design for
explicitness and control admit a wide range of policies.

We believe \proto's design offers a promising path for more closely integrating
measurement into protocol design. First, via \PRREF{cost}, \proto is designed so
that most of the actors simply provide small bits of information---leaving the
consumer to bear the analysis costs.  Second, the primitives are independent and
therefore can be implemented and deployed without a burdensome amount of
unwanted complexity. For instance, the end-to-end timing primitive
(\xref{sec:prim:e2e:time}) can be adopted without the primitive that provides
arrival information (\xref{sec:prim:e2e:arr}). The history of protocol
transitions shows that such functional independence is crucial to adoption.
Technologies that require coordinated deployment by a multitude of actors to
gain any benefit tend to be difficult to deploy (e.g., IPv6~\cite{CAZ+14},
ECN~\cite{KuNeTr13}). Third, as we argue in \xref{sec:practical:overhead},
\proto's overhead is low enough that it does not present a barrier to entry.
Finally, \proto does not need ubiquitous deployment to provide benefit.  This
manifests in a number of ways: ($i$) endpoints can leverage the end-to-end
primitives without any assistance from intermediate routers, ($ii$) endpoints
can derive benefits from \proto if only some of their peers support the facility
and ($iii$) incomplete information---such as might come from only some routers
along a path supporting \proto---is still better than the current void of
information.

We also note that deployment of new technologies is easier within certain
subsets of the Internet. For instance, within a homogeneous network under
unified administration, such as many data centers, fairly radical changes to the
end points, routers, and protocols can be undertaken. Data center networks and
virtualized, software-defined networks increasingly rely on measurements to
operate efficiently. Uptake within such specialized networks may drive
implementation, and in turn, availability of
\proto hop-specific and path accumulating primitives on the Internet
at large. 

Finally, we note that as new technologies emerge they inevitably change the
surrounding ecosystem.  In particular, adoption of a facility like \proto may be
driven by the loss of functionality as our protocols evolve. For example, the
passive timing and arrival inference possible with traditional protocols like
TCP are no longer available with protocols that radiate less information by
design, such as QUIC~\cite{quic}, rendering traditional techniques useless.

\subsection{Adversaries}
\label{sec:practical:advert}

A final practical consideration involves coping with adversaries
that aim to provide bogus information within \proto.  Since we
piggyback \proto on normal transactions, we get significant
protection from blind, off-path attacks by ensuring we do not
consider \proto information from invalid packets.  That is, if a
protocol would naturally discard a segment---e.g., a TCP segment not
within the current window---the \proto information should similarly
be discarded.  This leaves adversaries that are either legitimately
involved in the communication or that can actively modify packets as
they traverse the network.  For instance, an ISP providing
performance information may decide to report a queue length that is
smaller than the actual queue length to hide the presence of
bufferbloat in their network.  Or, consider $ISP_1$ that wishes to
make a competitor, $ISP_2$ look bad.  $ISP_1$ can increase the queue
length reported by $ISP_2$ in traffic as it passes through $ISP_1$.

One way to deal with this situation is to cryptographically sign
and/or encrypt \proto information.  Encryption runs counter to
\PRREF{visible} which dictates visibility into the \proto information.  While
signatures are possible, they are not foolproof and will likely end up
burdensome.  First, signatures do not help with the case where a legitimate
information source simply provides bogus data.  Second, soundly establishing
trust in the huge number of keys associated with the peers of a host or network
will be a significant undertaking in the best case.

Therefore, we intend to deal with adversaries using \proto's data collection,
statistics and discrepancy detection. For instance---to continue the example
from above---if we only observe $ISP_2$ advertising a large queue when the
traffic also traverses $ISP_1$ then we can start to treat these data points with
some skepticism. Similarly, finding queue length advertisements along a path
that suggest a significantly different end-to-end latency than we measure via
the timing tuple we can start to view the information speculatively.

\section{Conclusions and Future Work}
\label{sec:concl}

The importance of network measurement continues to grow to cope with the complex
and tangled mess that is the modern Internet. Researchers and operators cleverly
leverage various artifacts in network technologies to assess the network and in
turn be more efficient, improve performance, and better understand the
Internet's behavior and operation.  We assert that the time has come to think
about measurement as a first-class piece of the network architecture, rather
than embodied in a series of hacks. We offer \proto as a first step toward this
goal.  The primitives we define are intentionally not complicated, and embody a
set of principles formed from operational and research experience. The small
bits of \textit{explicit} information in \proto can be combined to form the
basis of sound network assessment, which is too often the opposite of the
current state of making inferences from whatever information can be gleaned from
error-prone and convoluted techniques. While admittedly modest, our hope is to
advance a process that we believe needs to be undertaken for the continued
evolution of the Internet.

We leave protocol definition and implementation details of each primitive as
future work. However, we note that work is underway within the IETF~\cite{plus}
to define a new layer in the protocol stack into which \proto can be slotted.
Further, the advent of programmable networks, e.g.,
OpenFlow~\cite{mckeown2008openflow} and P4~\cite{bosshart2014p4}, may provide a
ready avenue to rapidly deploy and test IPIM.

\noindent \textbf{Acknowledgments.} We thank the anonymous reviewers for
invaluable feedback. This work supported in part by National Science Foundation
(NSF) grants CNS-1213155, CNS-1213157 and CNS-1564329; the European Commission
(EC) under grant agreement 
H2020-688421; and the Swiss State Secretariat for Education, Research, and
Innovation under contract 15.0268. This work represents the position of the
authors and not of the NSF, EC, or the Swiss or U.S.\ governments.

{ \balance
{
\bibliographystyle{abbrv}
\bibliography{mip,mallman}


}
}

\end{document}